\documentclass[12pt]{revtex4}
\usepackage{float}
\usepackage{pslatex}
\usepackage[T1]{fontenc}
\usepackage{epsf}

\begin{document}

\title{Electrical and optical dipole polarizabilities of two lowest triplet
states of the beryllium atom}
\author{Tymoteusz Mrozek, Krzysztof Strasburger}
\affiliation{Department of Physical and Quantum Chemistry \\
Faculty of Chemistry \\
Wroc{\l}aw University of Science and Technology \\
Wybrze\.{z}e Wyspia\'{n}skiego 27, 50-370 Wroc{\l}aw, Poland}
\date{\today}

\begin{abstract}
High-accuracy static and dynamic polarizabilities of the $2^3P_u$ and
$^{3}S_{g}$ states of the neutral beryllium atom are reported.
The calculations employed explicitly correlated Gaussian (ECG) basis sets,
and first-order wave functions were obtained by minimization of the Hylleraas
functional. The final basis sets used for the response-property calculations
contained c.a. 7000-8000 functions, depending on the polarizability
component and the chosen symmetry. The results cover frequencies from the
static limit, \(\omega = 0\), up to the third resonance for the
\(2^{3}P_{u}\) state and up to the second resonance for the
\(3^{3}S_{g}\) state. To the best of our knowledge, these are the
first calculations of dynamic and static polarizabilities for these
excited states of the beryllium atom using explicitly correlated
basis sets.
\end{abstract}

\maketitle

\section{Introduction}

Neutral beryllium remains a strict benchmark for the atomic theory,
as a four-electron atom, for which highly accurate calculations are
feasible.To date, the research for this this atom was focused primarily on calculating
the energy levels, and their relativistic splittings,
for the ground and excited states \cite{Komasa1995BeECG,Bubin2009BeSingletP,Sharkey2014BeSingletTripletD,Kedziorski2020AtomicFineStructureECG,Puchalski2021BeFineHyperfineSplitting,Stanke2022Be3PFineStructure}.
Modern studies account for finite nuclear mass effects,
relativistic corrections, and quantum electrodynamic (QED)
contributions.
\cite{Pachucki2004BeRelQED,Stanke2009BeSingletSFiniteMass,Puchalski2013BeSingletQED,Puchalski2014BeIsotopeShift,Cook2018BeQEDExperiment,Stanke2019BeSingletPRydberg,Kedziorski2020AtomicFineStructureECG,Nasiri2021BeBenchmarkSpectraOscillatorStrengths,Puchalski2021BeTripletPHyperfineQuadrupole,Puchalski2021BeFineHyperfineSplitting,Stanke2021BeSingletDRydberg,Stanke2022BeTripletDRydberg,Stanke2022Be3PFineStructure,Nasiri2025BeTripletTransitions,Stanke2026BeSingletDFiniteMass,Rzhevskii2026BeLowestTripletTransitionEnergies}.

Accurate description of the ground and excited states provides a foundation
for the studies of response properties, such as polarizabilities.
Although the response properties of the ground state have been thoroughly
investigated
\cite{Pluta1992BeElectricProperties,Tunega1997BeStaticDipolePolarizability,
Komasa2001BeDipoleQuadrupolePolarizabilities,Cheng2013BeTuneOutWavelengths,Porsev2006BeGroundPolarizabilities,Singh2013BeGroundPolarizabilityRCC,Strasburger2014BePolarizabilityECG},
one of the least explored topics for the beryllium atom remains the
high-accuracy description of response properties in excited states.
In particular, no such
calculations have been performed for excited states of neutral beryllium
using an explicitly correlated basis set.
Only scarce results obtained with other methods are reported in the literature.

Early studies of excited-state electric properties were performed
in the 1990s, with the electronic-structure methods suitable for the
computational resources available at that time.
Among them, the static polarizabilities were calculated with a variant of
the finite-field method \cite{Themelis1995BeExcitedValencePolarizabilities}.
The time dependent gauge invariant (TDGI) method was employed
for the dynamic polarizabilities
\cite{Begue1998Be3SPolarizabilities,Begue1998Be2PPolarizabilities}.
In this method, the basis for the first-order wave function is
composed of the unperturbed wave function multiplied by a polynomial (gauge),
so-called spectroscopic (selected eigenvectors of the
Hamiltonian) and quasi-spectral states which are
additional terms, selected to take account for the
incompleteness of this representation, including part of the
continuum contribution.  

The Green function formalism has been applied for the calculation of the
polarizabilities of various atoms \cite{Chernov2005DynamicPolarizabilitiesBe},
but unfortunately without providing the results for the beryllium triplet
states -- although they have been computed, according to the article's text.
In 2010, Mitroy has obtained accurate static polarizabilities of the ground
state and the $2^{3}P_{u}$ state, with extensive configuration interaction
(CI) wave functions for the valence electrons, and the $1s^2$ core represented in
semi-empirical way\cite{Mitroy2010BeClockPolarizabilities}. In this approach,
the accuracy is mainly
determined by the completeness of the configuration space and by the
adopted treatment of the atomic core and core--valence correlation. The
use of an effective semi-empirical core-polarization potential allows
to reduce the computational cost, while the valence excitation space
is represented efficiently, but at the same time introduces a model
dependence that is absent in all-electron treatment.

In the more recent research from 2021 and 2023, methods were used that
take account of the relativistic effects, within either fully relativistic
multiconfiguration Dirac-Hartree-Fock (MCDHF) or relativistic
configuration-interaction plus core-polarization (RCICP) methods.
In the MCDHF calculation by Dong \cite{Dong2021BePolarizabilityMCDHF}, all four electrons are treated
explicitly, and the static response is determined by solving
an inhomogeneous equation derived from the sum-over-states relation.
In contrast, in the RCICP calculation \cite{Wu2023BeOpticalLatticeClocks}, the frozen Be$^{2+}$ core is
represented by effective potential in the Hamiltonian for the valence electrons.
This Hamiltonian is then diagonalized in two-electron configuration space,
and the polarizability is obtained with the sum-over-states method.
In both cases, relativistic effects are not corrections to the nonrelativistic
polarizabilities, therefore the results are not directly comparable with those
stemming from nonrelativistic calculations.

In the present work, the static and dynamic polarizabilities of two lowest
\(2^3P_{u}\) and \(3^3S_{g}\) triplet states of the beryllium atom
are calculated with nonrelativistic Hamiltonian, for inifinite-mass
nucleus. The response-function formalism is employed, in combination with
ECG basis sets, to provide benchmark data for these states. The application
of ECG functions furnished many most successful studies for few-electron
systems in the history of computational quantum chemistry \cite{RevModPhys.85.693}.
For the response problem, the first-order wave function is determined by
minimization of the Hylleraas functional at selected frequencies,
allowing the basis to adapt directly to the part of the spectrum relevant
for a given value of \(\omega\). The main advantage of this procedure is
the highly flexible, explicitly correlated representation of both the
reference and response wave functions.

Atomic units are used for all results.

\section{Methods}

In the following text, the ${\bf r}$ vector without particle index comprises
spatial coordinates of all electrons, the ${\bf x}$ vector -- their space-
and spin-coordinates, while the indexed vectors points at only one
electron. Atomic units are used, the reduced Planck constant is however
explicitly given in $\hbar\omega$ expressions, in order to make respective
energies clearly distinct from the angular frequencies.

Consider a system (atom, molecule) described with time-independent
Hamiltonian $\hat{H}^{(0)}$, with the eigenfunctions
fulfilling the stationary Schr\"{o}dinger equation

\begin{equation}
\hat{H}^{(0)}({\bf r})\Psi_K^{(0)}({\bf x})=E_K^{(0)}\Psi_K^{(0)}({\bf x}).
\label{SE-stat}
\end{equation}

The frequency-dependent dipole polarizability of this system is calculated
within the framework of time-dependent perturbation theory. The interaction
of the electrons with periodic external potential, such as the electric field
component of light, is described by the operator

\begin{equation}
\hat{V}({\bf r},t)=\hat{V}({\bf r})\cos(\omega t),
\label{periodic-pert}
\end{equation}

where $\omega$ is the circular frequency of the perturbation.
In the case of vanishing $\omega$, the static (electrical)
polarizability is obtained as the second derivative of the energy with
respect to the field intensity $F$, and related to the second order energy
correction, as $\alpha=-2E^{(2)}$. For time-dependent Hamiltonian, the energy
is no longer a constant of motion, therefore a generalization of the
stationary Schr\"{o}dinger equation is formulated, containing auxiliary
property called quasienergy \cite{Helgaker2012MolecularProperties}. This formalism is briefly
recapitulated here. The time-dependent wave function is written as

\begin{equation}
\Phi({\bf x},t)=\Psi({\bf x},t)e^{-\frac{i}{\hbar}\phi(t)}.
\end{equation}

For stationary states $\Psi_K^{(0)}$, the phase factor contains the function
$\phi(t)=E_K^{(0)}t+\chi$. For other states, $\phi(t)$ is required to be
real-valued.
Insertion of $\Phi({\bf x},t)$ into the time-dependent Schr\"{o}dinger equation,
then elimination of the phase factor, and multiplication by $\Psi^*({\bf x},t)$
yields, for normalized $\Psi({\bf x},t)$, the quasienergy

\begin{equation}
Q(t)=\frac{\partial\phi(t)}{\partial t}=
\langle\Psi({\bf x},t)\left|\hat{H}({\bf r},t)-i\hbar\frac{\partial}{\partial t}\right|\Psi({\bf x},t)\rangle
\end{equation}

which is a non-physical, real-valued, time-dependent function
that becomes the energy for $\omega=0$.
As intermediate normalization of the wave function is usually assumed in the
perturbation theory,
$\langle\Psi({\bf x},t)|\Psi^{(0)}({\bf x})\rangle=1$, the positive,
time-dependent normalization function $c(t)$ is defined. Insertion of
$\Phi({\bf x},t)=c(t)\Psi_({\bf x},t)e^{-\frac{i}{\hbar}\phi(t)}$
into the time-dependent Schr\"{o}dinger equation, leads to

\begin{equation}
\left[\hat{H}({\bf r},t)-i\hbar\frac{\partial}{\partial t}\right]\Psi({\bf x},t)=\eta(t)\Psi({\bf x},t).
\label{qe-td2}
\end{equation}

where

\begin{equation}
\eta(t)=Q(t)+i\hbar\frac{\partial\ln{c(t)}}{\partial t}
\end{equation}

may be a complex function of time, whose real part is the quasienergy, and
the imaginary part is related to the time evolution of the amplitude
of $\Psi({\bf x},t)$. It is convenient to decompose the operator given by
Eq. (\ref{periodic-pert}) to two complex components:
$\hat{V}_-({\bf r})e^{-i\omega t}$, and
$\hat{V}_+({\bf r})e^{i\omega t}$, and treat them formally as
independent perturbations, while obviously

\begin{equation}
\hat{V}_-({\bf r})=\hat{V}_+({\bf r})=\frac{\hat{V}({\bf r})}{2}.
\end{equation}

The Hamiltonian is written with double perturbation

\begin{equation}
\hat{H}({\bf r},t)=\hat{H}^{(0)}({\bf r})+\lambda_-\hat{V}_-({\bf r})e^{-i\omega t}+\lambda_+\hat{V}_+({\bf r})e^{(\omega t},
\end{equation}

and the solutions of Eq. (\ref{qe-td2}) as perturbative series'

\begin{equation}
\Psi({\bf x},t)=\Psi^{(0)}({\bf x})+
\lambda_-\Psi_{-}^{(1)}({\bf x})e^{-i\omega t}+
\lambda_+\Psi_{+}^{(1)}({\bf x})e^{i\omega t}+\ldots,
\label{wfn-pert}
\end{equation}

\begin{equation}
\eta(t)=E^{(0)}+\lambda_-\eta_-^{(1)}+\lambda_+\eta_+^{(1)}+
\lambda_-^2\eta_{--}^{(2)}+\lambda_-\lambda_+\eta_{-+}^{(2)}
+\lambda_+\lambda_-\eta_{+-}^{(2)}+\lambda_+^2\eta_{++}^{(2)}+\ldots.
\label{eta-pert}
\end{equation}

Both formal perturbation parameters, $\lambda_-$ and $\lambda_+$, are
interpreted as the amplitude of the electric component of the
electromagnetic field. Physical meaning is
attributed to the derivatives of the quasienergy averaged over
the period of the time-dependent potential ($T=\frac{2\pi}{\omega}$),
with respect to the external potential's parameters. The lowest order
non-oscillating terms are identified in the second-order calculus:
$\eta_{-+}^{(2)}=Q_{-+}^{(2)}=\langle\Psi_K^{(0)}({\bf x})|\hat{V}_-({\bf r})|\Psi_{+}^{(1)}({\bf x})\rangle$, and
$\eta_{+-}^{(2)}=Q_{+-}^{(2)}=\langle\Psi_K^{(0)}({\bf x})|\hat{V}_+({\bf r})|\Psi_{-}^{(1)}({\bf x})\rangle$.
The frequency-dependent polarizability is equal
\begin{equation}
\alpha(\omega)=-(Q_{-+}^{(2)}+Q_{+-}^{(2)})
\end{equation}

Assuming the $K^{th}$ unperturbed state is nondegenerate with any other state
that may contribute to the perturbative corrections, due to symmetry
constraints, the first-order wave functions $\Psi_{K\pm}^{(1)}$ may be
calculated either with the sum-over-states
(SOS) method, or by minimization of proper second-order Hylleraas functional

\begin{equation}
J_{K\pm}^{(2)}[\Psi_1]=\langle\Psi_1|\hat{H}^{(0)}-E_K^{(0)}\pm\hbar\omega|\Psi_1\rangle
+\langle\Psi_1|\hat{V}_{\pm}|\Psi_K^{(0)}\rangle
   +\langle\Psi_K^{(0)}|\hat{V}_{\pm}|\Psi_1\rangle.
\end{equation}

In the present work, the wave functions are expressed as linear combinations
of non-orthogonal basis functions, whose non-linear parameters are optimized to
minimize either the energy of given state or the desired second order 
correction to the quasienergy. Therefore the second method is preferred,
which has been proven successful in former calculations carried out for the
ground state of beryllium \cite{Komasa2001BeDipoleQuadrupolePolarizabilities}.
If $\Psi_K^{(0)}$ is an excited
state, then combined approach is necessary, at least at the stage of the
optimization of basis functions' parameters. The wave functions of states with
the energies lower than $E_K^{(0)}+\hbar\omega$
have to be projected out of the space of functions, in which the Hylleraas
functional is minimized \cite{Sellers}. Missing contributions of these states
are calculated explicitly as

\begin{equation}
Q_{I\pm}^{(2)}=-\frac{|\langle\Psi_K^{(0)}|V|\Psi_I^{(0)}\rangle|^2}{E_I^{(0)}-E_K^{(0)}\pm\hbar\omega}
\label{SOScontr}
\end{equation}

and added to the Hylleraas functional. Without this projection, variational
collapse in the basis optimization algorithm was observed, with the minimized
$J_{K\pm}^{(2)}$ functional falling down to minus infinity. Explanation for
such situations is based on the equivalence of the Hylleraas functional with
the sum over states. The contribution of the states, whose
energies are lower than that of the reference state, is positive, as long as
the denominator in Eq. \ref{SOScontr} remains negative.
If the basis is degraded to the point where the representation
of such state in the Hamiltonian matrix becomes very inaccurate, such that
$E_I^{(0)}\pm\hbar\omega$ raises above $E_K^{(0)}$, then the denominator in
the perturbative expression changes the sign, and the contribution to
$Q_{I\pm}^{(2)}$ becomes negative. Further optimization drives the basis
towards production of a false pole.

The nonrelativistic Hamiltonian for the $n$-electron atom with infinite
nuclear mass is

\begin{equation}
\hat{H}^{(0)}=-\frac{1}{2}\sum_{i=1}^{n}\nabla_{i}^{2}
-\sum_{i=1}^{n}\frac{Z}{r_{i}}+\sum_{i<j}^{n}\frac{1}{r_{ij}}
\end{equation}

with $n=4$ and $Z=4$ for beryllium. The perturbation is given by

\begin{equation}
\hat{V}({\bf r}_1,\ldots,{\bf r}_n)=\sum_{i=1}^n{\bf F}\cdot\bf{r}_i
\end{equation}

where ${\bf F}$ is the electric field intensity vector, which gets replaced
in the calculations by the unit vector, and the (scalar) intensity is
treated as the formal perturbation parameter.

All wave functions are expressed as
linear combinations of properly symmetrized explicitly correlated Gaussian
lobe functions.
The primitive $n$-electron ECG lobe function is defined as

\begin{equation}
\chi_I({\bf r}_1,\ldots,{\bf r}_n)=
\exp\left[-\sum_{i=1}^{n}\alpha_{i,I}\left(\mathbf{r}_{i}-\mathbf{R}_{i,I}\right)^2
-\sum_{i>j}^{n}\beta_{ij,I}\left(\mathbf{r}_{i}-\mathbf{r}_{j}\right)^{2}\right]
\label{ecglobe}
\end{equation}

where $\alpha_{i,I}$, $\beta_{ij,I}$, and $\mathbf{R}_{i,I}$ are nonlinear
variational parameters. The displacement vectors ${\bf R}_{i,I}$ generate
the lobe character of the basis. In general, a single primitive function
defined by Eq.~\ref{ecglobe} is not an eigenfunction of the $\hat{L}^{2}$
operator, except of the nucleus-centered $L=0$ case. Therefore, the desired
angular symmetry is imposed approximately by point-group projection rather
than by multiplying the Gaussian functions by explicit angular polynomials.

Any-order wave function is expressed as

\begin{equation}
\Psi_K({\bf x}_1,\ldots,{\bf x}_n)=\sum_{I=1}^Nc_I
\hat{A} \left[ \Theta_{S,M_S}\hat{P}_{G,\Gamma}
\chi_I({\bf r}_1,\ldots,{\bf r}_n) \right]
\label{wave-function}
\end{equation}
with the perturbative corrections sharing the same spin function
$\Theta_{S,M_{S}}$ with the unperturbed $\Psi_K^{(0)}$, as neither the
Hamiltonian (\ref{SE-stat}) nor the perturbation (\ref{periodic-pert})
depend on the spins. $\hat{A}$ is the antisymmetrizer that cares for proper
permutational symmetry of the fermionic wave function, and $\hat{P}_{G,\Gamma}$
is the spatial symmetry projector proper for one-dimensional irreducible
representation $\Gamma$ of finite point group $G$.
\begin{equation}
\hat{P}_{G,\Gamma}=\sum_{g \in G}p_{\Gamma}(g)\hat{R}(g),
\end{equation}
where $p_{\Gamma}(g)$ is the character of the operation $g$, and $\hat{R}(g)$
is the symmetry operation that acts on all displacement vectors
${\bf R}_{i,I}$.

The linear coefficients for unperturbed states, with given set of the
nonlinear parameters (the latter are optimized with the conjugate gradient
method) are obtained from the generalized matrix eigenvalue problem
\begin{equation}
{\bf H}^{(0)}{\bf c}^{(0)} = E_{0}{\bf S}^{(0)}{\bf c}^{(0)}.
\end{equation}

The energy is minimized variationally in the space spanned by the
symmetry-projected basis functions. The expectation value
$\langle\hat{L}^{2}\rangle$ is expected to converge towards the value of
$L(L+1)$ proper for given unperturbed state, and is used as a diagnostic
of angular purity. This is because the point-group projector eliminates
many lower-$L$ contaminants, whereas residual higher-$L$ admixtures give
a positive contribution to the variational energy. Thus, as the basis is
enlarged and optimized, the energy decreases and $\langle\hat{L}^{2}\rangle$
approaches the exact value.
Concerning the spatial symmetry projectors, they are chosen individually
for each state, with a reasonable compromise in mind, between the size
of the projected basis set, and the number of operations needed to complete
the calculation of the matrix elements.

\section{RESULTS AND DISCUSSION}

\subsection{Construction of the wave functions} 

In the calculations of the polarizabilities, or other response properties,
by the minimization of the Hylleraas functional, the wave function of the
reference (unperturbed) state is expected to be as accurate as possible.
The energy of both the $2^3P_u$ and $^3S_g$ states was within
1 $\mu$hartree above the variational limit. The wave function of the former
was calculated in the $A_u$ representation of the $C_i$ symmetry, with all
${\bf R}_{i,I}$ vectors confined to the $z$-axis of the coordinate system.
The symmetry projector was
\begin{equation}
\hat{P}_{C_i,A_u}=\hat{E}-\hat{i},
\label{CiAu}
\end{equation}
The latter state's wave function is spherically symmetric, so no projection
of the basis functions was needed. Formally it was realized by the $C_1$ point
group, with all ${\bf R}_{i,I}$ vectors set to zero.

There are additional excited states of the Be atom, whose energies
fall below the $E_K^{(0)}+\hbar\omega$ limit for sufficiently high frequency
in the considered range. They are the lowest $^3P_g$, and $^3D_g$ triplets
for $2^3P_u$, and $3^3P_u$ for $^3S_g$ state.
For practical reasons, their contributions to the polarizability were
universally calculated with the SOS method, and their basis functions added to
the set intended for the Hylleraas functional. They manifest themselves as
poles of the $\alpha(\omega)$ function, at the resonance frequencies.

The resonance regions are particularly sensitive to
the variations of the state energies. As $\hbar\omega$ approaches the
transition energy $E_I^{(0)}-E_K^{(0)}$, the corresponding denominator
in the perturbative expression (Eq. \ref{SOScontr}) becomes very small.
Consequently, even small change in $E_I^{(0)}$ is strongly amplified  in the
calculated polarizability, close to a resonance. Therefore the wave functions
of these states were optimized independently on the first order functions.

\begin{table}[h]
\centering
\caption{Nonrelativistic energies of the selected triplet states
of Be in the infinite-nuclear-mass limit.}
\label{tab:selected_states}

\begin{tabular}{lccllc}
\hline
State & $N$ & Energy [$E_h$] & $\langle L^2\rangle$ & Ref. energy [$E_h$] &  Error [$\mu E_h$]\\
\hline
 $2\,^3P_u$   & 2263 & $-14.56724368$ & $2.00000004$ & $-14.567244232(8)^a$  & $0.55$ \\
 $2\,^3P_u^c$ & 6907 & $-14.56724373$ & $2.00000003$ & $-14.567244232(8)^a$ & $0.50$ \\
 $3\,^3P_u^c$ & 6907 & $-14.39896587$ & $2.00000206$ & $-14.3989686969(88)^b$ & $2.83$ $$ \\
 $3\,{}^3S_g$ & 3201 & $-14.43006702$ & $0$ & $-14.430067678(7)^a$   & $0.66$ \\
 $3\,{}^3D_g (M_L=0)$ & 2116 & $-14.3846280$ & $6.00002133$ & $-14.384634616(7)^a$  & $6.62$ \\
 $3\,{}^3D_g (|M_L|=2)$ & 2408 & $-14.38463415$ & $6.00000104$ & $-14.384634616(7)^a$ & $0.47$ \\
 $2p^2,{}^3P_g (|M_L|=0)$ & 2422 & $-14.39545300$ & $2.00000026$ & $-14.395453745(4)^a$   & $0.75$ \\
\hline

\multicolumn{6}{l}{$^a$ Ref. \cite{Puchalski2021BeFineHyperfineSplitting},
$^b$ Ref. \cite{Nasiri2025BeTripletTransitions,Rzhevskii2026BeLowestTripletTransitionEnergies},
$^c$ Combined basis of all functions with proper symmetry}
\end{tabular}
\end{table}

Considering the triply degenerate $^3P_u$ state, the calculations
were carried out for the ($L=1$, $M_L=0$) case, for which typical orientation
of the coordinate axes was assumed, with $xy$-plane as the nodal surface.
With the electric field acting
along the $z$-axis, the wave functions of the $^3S_g$ and the lowest $^3D_g$
($L=2$, $M_L=0$) states are needed for the calculation of the $\alpha_{zz}$
component of the polarizability tensor. The former is discussed above, while
it is quite difficult to describe the wave function of the latter with the lobe
functions. The non-flat nodal surface proper for this wave function
cannot be produced by symmetry projectors of any one-dimensional representation
of any point group. As a compromise, the fully symmetrical $A_{1g}$ representation
of the $D_{4h}$ point group was employed, with freely floating ${\bf R}_{i,I}$
vectors. The nodal surface had to appear spontaneously. The basis was
large enough to provide at least a rough representation of the $^3S_g$
state, which appears in the same representation in the Hamiltonian
spectrum, so that the collapse to this state was avoided while optimizing
the nonlinear parameters of the basis functions.
This approach to the problem appeared to be successful, but at the price of
significantly slower convergence with the basis set size, than for the
state with $|M_L|=2$. The final energy of this state was almost
7 $\mu$hartree above the variational limit (table \ref{tab:selected_states}).
Remaining part of the first-order function for this polarizability component 
was obtained in the $A_g$ representation of the $C_i$ group, with the
projector
\begin{equation}
\hat{P}_{C_i,A_g}=\hat{E}+\hat{i},
\label{CiAg}
\end{equation}
and all ${\bf R}_{i,I}$ vectors confined to the $z$ axis.

The $\alpha_{xx}$ and $\alpha_{yy}$ components are degenerate, so it is
sufficient to calculate only one. The $^3P_g$ and $^3D_g$ states contribute
to these components of the polarizability. The most convenient solution of
the computational problem was to rotate the wave function of the $^3P_u$ state,
by setting the ${\bf R}_{i,I}$ vectors along the $x$ axis, thus making it
proper for $|M_L|=1$, act with the
electric field in the $y$ direction, and express the wave functions of the
lowest $^3P_g$ and $^3D_g$ states in the $D_{4h}$ symmetry with the main
rotation axis fixed to the $z$-axis, in the $A_{2g}$
\begin{equation}
\hat{P}_{D_{4h},A_{2g}}=\hat{E}+\hat{C}_4^1+\hat{C}_2+\hat{C}_4^3
-\hat{C}'_{2,1}-\hat{C}'_{2,2}-\hat{C}''_{2,1}-\hat{C}''_{2,2}
+\hat{S}_4^1+\hat{i}+\hat{S}_4^3+\hat{\sigma}_h
-\hat{\sigma}_{v,1}-\hat{\sigma}_{v,2}-\hat{\sigma}_{d,1}-\hat{\sigma}_{d,2},
\end{equation}
and $B_{2g}$
\begin{equation}
\hat{P}_{D_{4h},B_{2g}}=\hat{E}-\hat{C}_4^1+\hat{C}_2-\hat{C}_4^3
-\hat{C}'_{2,1}-\hat{C}'_{2,2}+\hat{C}''_{2,1}+\hat{C}''_{2,2}
-\hat{S}_4^1+\hat{i}-\hat{S}_4^3+\hat{\sigma}_h
-\hat{\sigma}_{v,1}-\hat{\sigma}_{v,2}+\hat{\sigma}_{d,1}+\hat{\sigma}_{d,2}
\end{equation}
representations, respectively. In these cases, the symmetry
projectors produce
correct nodal surfaces of the wave functions, making it possible to obtain the
energies accurate within 1 $\mu$hartree with basis sets of reasonable size.
The $C_{2v}$ point group was employed for
the basis for remaining part of the first-order wave function, calculated with
the Hylleraas functional. The basis functions were confined to the $xy$ plane,
and the projector was proper for the $A_2$ representation,
\begin{equation}
\hat{P}_{C_{2v},A_2}=\hat{E}+\hat{C}_{2}-\hat{\sigma}_{xz}-\hat{\sigma}_{yz}.
\end{equation}

\begin{table}
\centering
\caption{Convergence of the polarizability components of the $2\,^3P_u$ state,
for selected frequencies (given in parentheses)}
\label{table-convergence-1}
\begin{tabular}{lcccccc}
\hline
\multicolumn{7}{c}{The number of basis functions} \\
 & 200 & 600 & 800 & 1000 & 2000 & Final \\
\hline
$\alpha_{zz}(0)$    & 37.924 & 37.929 & 37.936 & 37.938 & 37.943 & 37.945 \\
$\alpha_{xx}=\alpha_{yy}(0)$    & 39.554 & 39.561 & 39.562 & 39.563 & 39.564 & 39.565 \\
$\alpha_0(0)$    & 39.010 & 39.017 & 39.020 & 39.022 & 39.024 & 39.025 \\
$\alpha_{zz}(0.1)$    & 60.089 & 60.095 & 60.102 & 60.106 & 60.112 & 60.113 \\
$\alpha_{xx}=\alpha_{yy}(0.1)$    & 56.211 & 56.219 & 56.220 & 56.222 & 56.223 & 56.224 \\
$\alpha_0(0.1)$ & 57.504 & 57.511 & 57.514 & 57.516 & 57.519 & 57.520 \\
$\alpha_{zz}(0.15)$    & -10.976 & -10.968 & -10.958 & -10.953 & -10.945 & -10.943 \\
$\alpha_{xx}=\alpha_{yy}(0.15)$    & 133.954 & 133.963 & 133.965 & 133.966 & 133.968 & 133.969 \\
$\alpha_0(0.15)$ & 85.644 & 85.653 & 85.657 & 85.660 & 85.663 & 85.665 \\
\hline
\end{tabular}
\end{table}

\begin{table}
\centering
\caption{Convergence of the polarizability of the $3^3S_g$ state, for
selected frequencies (given in parentheses)}
\label{table-convergence-2}

\begin{tabular}{lccccc}
\hline
& \multicolumn{5}{c}{The number of basis functions} \\
& 400 & 1200 & 1600 & 2000 & Final \\
\hline
$\alpha_0(0)$    & 1173.14& 1174.51  & 1174.72  & 1174.73  & 1174.82 \\
$\alpha_0(0.02)$ & 1993.23& 1995.06  & 1995.33  & 1995.39  & 1995.48 \\
$\alpha_0(0.04)$ & -1768.86& -1767.40 & -1767.12 & -1767.10 & -1766.93 \\
\hline
\end{tabular}

\end{table}

Concerning the polarizability of the $^3S_g$ state, all three compoments are
degenerate. Apart of the lowest $^3P_u$ state, the higher state with the same
symmetry was included in the SOS calculations. The $A_u$ representation of
the $C_i$ point group was used for both, as well as for the remaining part
of the first-order function, with ${\bf R}_{i,I}$ vectors placed on the
$z$ axis.

Obviously, the physically relevant polarizability of an atom in gas phase
is obtained by averaging
\begin{equation}
\alpha_0=\frac{1}{3}(\alpha_{xx}+\alpha_{yy}+\alpha_{zz})
\end{equation}

and its anisotropy is calculated for the $P$-symmetry state,
\begin{equation}
\Delta\alpha(\omega)=\frac{1}{3}\left[\alpha_{xx}(\omega)-\alpha_{zz}(\omega)\right].
\label{anisotr}
\end{equation}

Numerical values of the energies of all explicitly included states are
presented in Table \ref{tab:selected_states}, together with the final number
of used basis functions. The $3^3D_g$ state occurs twice, with different
energies, because the wave functions had to be built separately for $M_L=0$,
and $|M_L|=2$ (the latter is real-valued linear combination of complex
wave functions with $M_L=\pm 2$).

The basis sets for first-order wave functions were constructed over a range
of $\omega$, using similar scheme for adding new functions. Beginning at
$\omega=0$, the parameters of an initial set of functions were optimized
to minimize the Hylleraas functional. After completing this step, with
the convergence criterion of 1 mhartree per cycle, new value of omega was 
chosen. New set of functions was then added to the basis, and their
paramaters were optimized. This procedure
was repeated for consecutive values of $\omega$ until reaching the assumed
limit, which appeared close to first resonance frequency.
The functions contributing less than 1 nhartree
to the value of the functional were dropped from the basis.
Independent response bases were constructed for the
positive- and negative-frequency contributions. In final calculations, they
were concatenated together, and with the bases built for the states included
explicitly with the SOS method. The convergence of the polarizability
was sufficient to assume that four decimal digits of the results are
accurate (tables \ref{table-convergence-1} and \ref{table-convergence-2}).

\subsection{$2^3P_u$ state}

\begin{figure}[htbp]
    \centering
    \epsfbox{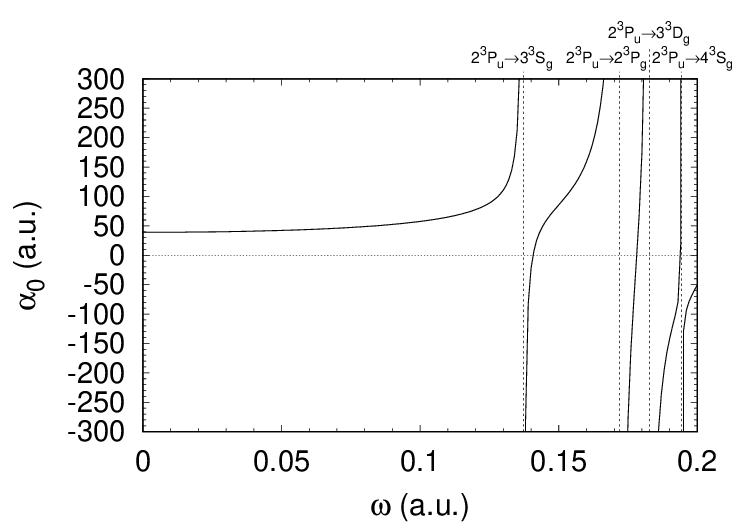}
    \caption{Frequency dependence of the scalar polarizability
    of the $2^3P_u$ state}
    \label{fig:2Palpha0}
\end{figure}

\begin{figure}[htbp]
    \centering
    \epsfbox{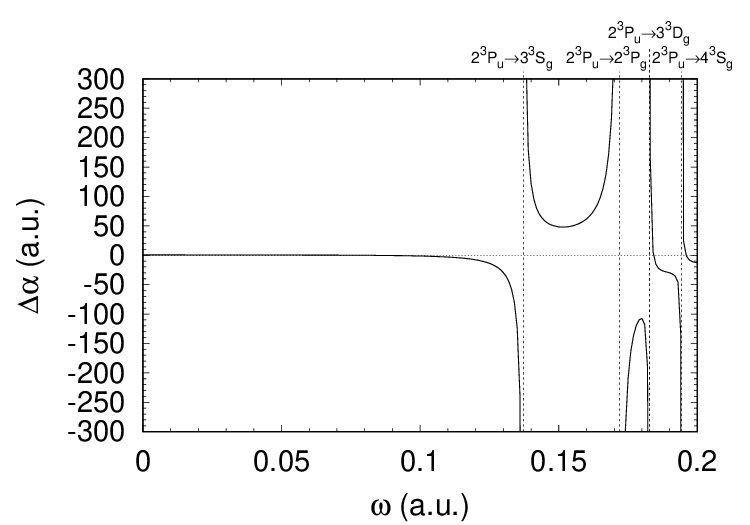}
    \caption{Frequency dependence of the polarizability anisotropy
    of the $2^3P_u$ state}
    \label{fig:2Pdeltaalpha}
\end{figure}

For the $2^3P_u$ state, the basis composed of 2263 ECG functions was
constructed first. The optimization of the response basis was
carried out at several frequencies, starting with 200 functions at
$\omega=0$ and extending this set for several, increasing frequencies,
selected to sample the low-frequency region and slightly above
first resonance. Thus, a reliable description of the response over
this frequency interval was maintained.
The final sets contained 2591 basis functions for the ``parallel''
$\alpha_{zz}$ component, and 3508 functions for the ``perpendicular''
$\alpha_{yy}$ one. They were combined with the bases given in table
\ref{tab:selected_states} -- the former with these for the $3^3D_g (M_L=0)$ and
$3^3S_g$ states, giving the total of 7908 functions for response
calculations, and the latter with these for the $2^3P_g$ and $3^3D_g$
($|M_L|=2$) states, giving 8338 functions. The resulting scalar polarizability
is presented in Fig.~\ref{fig:2Palpha0}, and its anisotropy in
Fig.~\ref{fig:2Pdeltaalpha},
for the $\omega$ range from 0 to $0.2$ a.u. The frequency dependence reveals
the poles that represent the transitions to excited states. Their positions
correspond with the eigenvalues of the Hamiltonian matrix in the basis
for first-order wave function. Beyond the frequency of $0.19$ a.u., this basis
does not contain functions optimized for higher excited states, therefore their
resulting energies are increasingly
inaccurate and so are the polarizability poles. For example, the first
resonance that appears in the results spontaneously, i.e. without explicit
addition of basis functions optimized for particular state, at 0.19428 a.u.,
may be attributed to the transition to the $4^3S_g$ state. The respective
eigenvalue of the $H$ matrix amounts to $-14.372962$ hartree (almost 0.5
mhartree above the best variational energy of $-14.3734537307$ hartree
\cite{Nasiri2025BeTripletTransitions}), and the $\langle L^2\rangle$
expectation value is as inaccurate as $0.0446$.

\begin{table}
\centering
\caption{The polarizability of the $2^3P_u$ state of Be
-- comparison with the literature data}
\label{tab:2P-polarizability-005}
\begin{tabular}{ccccccc} \hline
$\omega$ & $\alpha_{xx}=\alpha_{yy}$&$\alpha_{zz}$ & $\alpha_0$ & $\Delta\alpha$ & $\alpha_0^a$ & $\Delta\alpha^a$ \\ \hline
0.000 & $39.565$ & $37.944$ & $39.025$ & $0.5401$ & $36.08$ & $1.04$ \\
      &          &          &          &          & $39.04(28)^b$ & $0.541^b$ \\
      &          &          &          &          & $39.33^c$ & \\
      &          &          &          &          & $39.1121^d$ & \\
      &          &          &          &          & $39.249^e$ & \\
0.010 & $39.680$ & $38.069$ & $39.143$ & $0.5368$ & $36.19$ &  $1.06$ \\
0.020 & $40.027$ & $38.449$ & $39.501$ & $0.5262$ & $36.52$ &  $1.12$ \\
0.030 & $40.621$ & $39.103$ & $40.115$ & $0.5062$ & $37.09$ &  $1.23$ \\
0.040 & $41.486$ & $40.067$ & $41.013$ & $0.4730$ & $37.92$ &  $1.39$ \\
0.050 & $42.658$ & $41.401$ & $42.239$ & $0.4193$ & $39.06$ &  $1.63$ \\
0.060 & $44.193$ & $43.195$ & $43.860$ & $0.3327$ & $40.59$ &  $1.96$ \\
0.070 & $46.168$ & $45.596$ & $45.977$ & $0.1906$ & $42.62$ &  $2.43$ \\
0.080 & $48.700$ & $48.850$ & $48.750$ & $-0.050$ & $45.34$ &  $3.12$ \\
0.090 & $51.962$ & $53.395$ & $52.440$ & $-0.4776$ & $49.09$ &  $4.17$ \\
0.100 & $56.224$ & $60.113$ & $57.520$ & $-1.2965$ & $54.59$ & $5.89$ \\
0.110 & $61.923$ & $71.107$ & $64.984$ & $-3.061$ & $63.51$ & $9.07$ \\
0.120 & $69.823$ & $93.159$ & $77.602$ & $-7.779$ & $81.63$ & $16.49$ \\
0.130 & $81.389$ & $169.873$ & $110.883$ & $-29.495$ & $153.09$ & $50.08$ \\
0.132 & $84.370$ & $219.161$ & $129.300$ & $-44.930$ & $211.07$ & $78.37$ \\
0.135 & $89.419$ & $459.789$ & $212.875$ & $-123.456$ & $800.74$ & $372.57$ \\
0.138 & $95.318$ &$-1041.448$ & $-283.605$ & $378.922$ & $-89.12$ & $-146.94$ \\
0.140 & $99.835$ & $-267.662$ & $-22.665$ & $122.499$ & $-18.91$ & $-78.43$ \\
0.145 & $113.895$ & $-60.590$ & $55.733$  & $58.162$ & $8.32$   & $-37.36$ \\
0.150 & $133.969$ & $-10.943$ & $85.665$  & $48.304$ & $42.24$  & $-25.81$ \\
0.160 & $220.756$ & $35.370$ & $158.960$ & $61.795$ & $65.43$  & $-16.68$ \\
0.170 & $1169.928$ & $85.131$ & $808.326$ & $361.600$ & $99.80$  & $-12.16$ \\
0.180 & $68.902$ & $391.568$ & $176.413$ & $-107.555$ & $250.35$ & $-4.21$ \\
0.182 & $1027.528$ & $1596.084$ & $1216.255$ & $189.518$ & $411.25$ & $0.99$ \\
0.184 & $-650.791$ & $-669.698$ & $-657.264$ & $6.302$ & $1494.52$ & $41.94$ \\
0.185 & $-420.152$ & $-374.699$ & $-405.065$ & $-15.151$ & $-1110.68$ & $-1111.06$ \\
0.187 & $-261.544$ & $-184.574$ & $-235.917$ & $-25.657$ & $-551.81$ & $-389.16$ \\
0.190 & $-171.431$ & $-81.924$ & $-141.640$ & $-29.836$ & $-30.07$ & $-142.29$ \\
\hline
\multicolumn{7}{l}{
$^a$ Ref. \cite{Begue1998Be2PPolarizabilities} unless stated otherwise,
$^b$\cite{Mitroy2010BeClockPolarizabilities},
$^c$ Ref. \cite{Themelis1995BeExcitedValencePolarizabilities}.}
\end{tabular}
\end{table}

The obtained polarizabilities are juxtaposed with the literature data, in table
\ref{tab:2P-polarizability-005}, focusing on the comparison with TDGI method
\cite{Begue1998Be2PPolarizabilities}, so that only the frequencies reported
in that work are selected. Full set of our results, used for preparation of
the plots, is available in the supplementary materials.
There is reasonable agreement only up to
the frequency slightly below the excitation to the $3^3S_g$ state, which
appears above 0.135 a.u. It is to be noted that the general idea of the TDGI
method is similar to the present work (in the sense of yielding variationally
bound polarizability), the underlying response space is however
constructed in fundamentally different way. The results' accuracy depends
on the quality of the reference state's wave function, but also on how reliably
the relevant intermediate spectrum is reproduced.
Agreement in the static and slowly varying
non-resonant frequency regions reflects mainly the overall quality of the
response description, whereas the discrepancies near resonances are much more
sensitive to differences in the calculated excitation energies and
transition matrix elements. Obviously, the resonance positions that are
determined by the energy separation between the reference and excited states,
are particularly sensitive to the quality of the corresponding wave functions,
and these require proper basis sets. This aspect becomes
particularly important for dynamic properties, since small inaccuracies
in the excitation energies have only a moderate influence on the final result
far from resonances, but may be strongly amplified when the frequency
approaches a pole. An interesting, seemingly qualitative difference,
is the lack of second pole in the TDGI polarizabilities, expected above
$0.171$ a.u., and corresponding to the excitation to the $^3P_g$ state.
It is likely that the pole frequency is shifted towards higher $\omega$ value,
proper rather
for the excitation to first $^3D_g$ state. Such errors in older works may arise
from limited size of the orbital basis sets, resulting in inaccurate
representation of the reference-state wave function, and the spectral states
as well. The highly converged energies obtained in the present
approach lead to more accurate resonance positions and, consequently, to
a more reliable frequency dependence of the polarizability in their
vicinities.

The anisotropy of the polarizability behaves completely differently than
in the TDGI results, pointing at apparent error in the process of calculation
of $\Delta\alpha$ in the latter.
The largest contribution to the polarizability stems from the $^3S_g$ state,
as $\omega$ approaches first resonance frequency. Therefore the $\alpha_{zz}$
component increases and exceeds $\alpha_{xx}$, for $\omega$ above 0.80 a.u.
(see table \ref{tab:2P-polarizability-005}). According to equation
(\ref{anisotr}), $\Delta\alpha$ should then become more and more negative,
and so are our results, up to the first resonance. To the contrary, opposite
sign of $\Delta\alpha$ is reported in the TDGI case
\cite{Begue1998Be2PPolarizabilities}. This inconsistency has been noticed
\cite{Chernov2005DynamicPolarizabilitiesBe}, but without reporting any
numerical results.

The static polarizability obtained by Themelis and Nikolaides \cite{Themelis1995BeExcitedValencePolarizabilities} in
finite-field-like calculations, seems overestimated by less than 1\%.
The most accurate result to date, including the polarizability anisotropy, has
been reported for this state by Mitroy \cite{Mitroy2010BeClockPolarizabilities}.
It agrees well with our result, within the estimated error margin. It was
obtained with a different computational methodology, and carefully
constructed CI wave function of the reference state. The scalar polarizability
is a little larger, but this difference may be related to the use of a
semi-empirical approximation for the beryllium core, while in the ECG
approach used in the present work, all electrons are treated explicitly.

\subsection{$^3S_g$ state}

The basis for the first-order function for the $^3S_g$ state was optimized
at several frequencies in the range below the first
resonance, corresponding to the $(3,^3S \rightarrow 3,^3P_u)$ transition.
The initial set was composed of 400 functions, and
the final response basis contained 2991 functions. It was concatenated
with  the basis for the $2^3P_u$ state, containing 2263 functions, and with the
set of 1023 functions describing the $3^3P_u$ state, optimized
in the presence of frozen basis for the $2^3P_u$ state. 

\begin{figure}[htbp]
    \centering
    \epsfbox{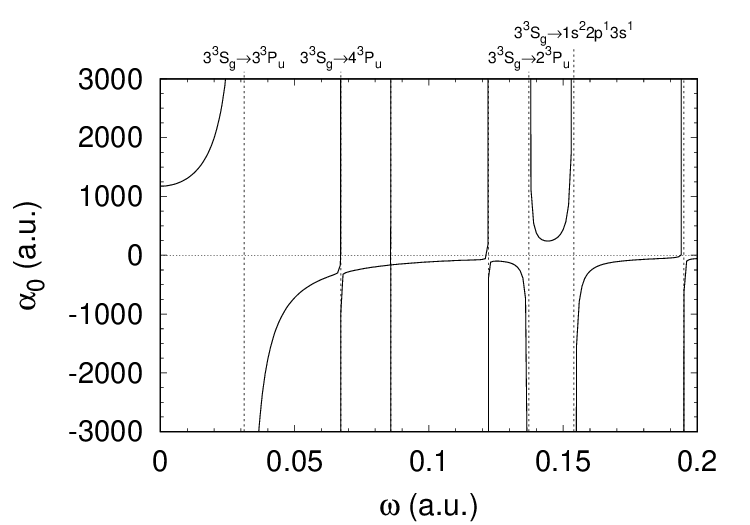}
    \caption{Frequency dependence of the polarizability
    of the $3^3S_g$ state}
    \label{fig:3Salpha0}
\end{figure}

Available literature data for this state are very limited. Compared with
the TDGI results\cite{Begue1998Be3SPolarizabilities}, the present polarizabilities listed in table
\ref{tab:3S-polarizability} are systematically lower
by few percent. Such a discrepancy is not alarming, when approaching a
resonance, but it occurs even in the low-frequency range. Both the present
method and TDGI are variational, therefore more accurate first-order wave
function is expected to yield a larger polarizability.
In order to make sure that no significant part of this function
remained underrepresented in the basis set, additional set of 630 ECG functions
was optimized quite roughly for the $4^3P_u$ state, and combined with
the basis meant for the first-order function, giving the total of 6907 basis
functions for this purpose. This extension of the basis
resulted in the static polarizability becoming larger by only 0.02 a.u.
(at 6$^{\text{th}}$ significant digit). Significant influence of the added set
on the result was observed only
in close vicinity of the frequency proper for the
$3^3S_g \rightarrow 4^3P_u$ transition. 

\begin{table}[h]
\centering
\caption{The polarizability of the $3\,{}^3S_g$ state of Be -- comparison
with the literature data}
\label{tab:3S-polarizability}
\begin{tabular}{ccc|ccc}
\hline
$\omega$ & $\alpha_0$ & $\alpha_0^{a}$ &
$\omega$ & $\alpha_0$ & $\alpha_0^{a}$ \\
\hline
0.0000 & 1174.82 & 1191.8 & 0.0250 & 3299.85  & 3206.5  \\
0.0025 & 1182.39 & 1199.1 & 0.0275 & 5345.10  & 4993.6  \\
0.0050 & 1205.71 & 1222.1 & 0.0295 & 11611.48 & 9704.0  \\
0.0075 & 1246.69 & 1262.6 & 0.0375 & -2552.21 & 22796.8 \\
0.0100 & 1309.03 & 1324.0 & 0.0400 & -1766.93 & -1902.4 \\
0.0125 & 1399.07 & 1412.5 & 0.0425 & -1329.47 & -1417.1 \\
0.0150 & 1527.63 & 1538.2 & 0.0450 & -1051.86 & -1113.7 \\
0.0175 & 1714.05 & 1719.3 & 0.0475 & -860.76  & -906.7  \\
0.0200 & 1995.48 & 1990.1 & 0.0500 & -721.66  & -756.9  \\
0.0225 & 2452.79 & 2423.8 & 0.0525 & -616.18  & -643.7  \\
\hline
\multicolumn{6}{l}{$^{a}$ Ref.~\cite{Begue1998Be3SPolarizabilities}.}
\end{tabular}
\end{table}

As our calculations are
converged to 4 significant digits, the differences between the
present results and those of old TDGI calculation using relatively limited
(12s4p3d1f) orbital basis and CI expansions of approximately
$2.5\times10^4$ determinants, are probably to be attributed to much
better wave function of the reference state in the basis of ECG functions.
The present results should therefore provide a more accurate
description of both the nonresonant response and the behaviour of the
polarizability in the vicinity of the $3^3S_g \rightarrow 3^3P_u$
resonance. With the energies of the $2^3P_u$ and $3^3P_u$ states
accurate within $0.5$ and $2.8$ $\mu$hartree, respectively,
and $4^3P_u$ represented fairly (off by 50 $\mu$hartree in comparison with
\cite{Nasiri2025BeTripletTransitions}) in the basis for first-order
wave function,
the polarizability calculations were carried out up to the frequency of
$0.2$ a.u., which is above the ionization limit. The results are presented in 
Fig.~\ref{fig:3Salpha0}. Above the 
$3^3S_g \rightarrow 4^3P_u$ transition frequency, for which quite a narrow
resonance is observed at $0.067166$ a.u., a forest of even narrower resonances 
would be expected, if the wave functions of respective excited $^3P_u$ states
would be represented in the response basis. This is not the case, therefore
there is only one lone pole below the ionization limit, at 0.0858 a.u.,
and an apparently false one above this limit, at 0.1221 a.u., where no
states exist at all. One may say that both occur at random frequencies.
However, the deexcitation ($3^3S_g\rightarrow 2^3P_u$) is described very
accurately, with the polarizability pole at 0.137177 a.u. This is by no
means a surprise, with the $2^3P_u$ state considered in the SOS manner. More
interesting is the pole at even higher frequency of 0.1539 a.u.
This pole may be attributed to the excitation to the resonance state
associated with the $1s^22p^13s^1$ electron configuration.
It appears spontaneously in the results, with the transition energy
less than 2 mhartree above the NIST-reported value of 0.15252 a.u. \cite{NIST},
and furnishes an example of a state represented fairly in the basis
constructed without this state in mind.

\section{Summary and conclusions}

High-quality static and dynamic polarizabilities of the beryllium atom
in the $2^3P_u$ and $3^3S_g$ excited triplet states are reported,
obtained with explicitly correlated Gaussian functions. The nonlinear
parameters of these functions were optimized within variation-perturbation
scheme involving minimization of the second-order Hylleraas functionals,
in the frequency range up to first resonance.
Concatenation with the sets constructed for wave functions of few additional
excited states, to which radiative dipole transitions are allowed, assured
accurate representation of low lying excitations, manifesting themselves by
polarizability poles.

Our results are expected to be significantly more accurate than those reported
in older literature. They remain in rather good agreement for low frequencies.
There are however some unexpected differences, like the polarizability of
the $3^3S_g$ state being smaller than in the TDGI calculation, that was
carried out with less accurate wave function of the reference
state. This difference brings to mind that any second-order property is
variationally bound in variation-perturbation methods, provided the
unperturbed wave function is exact.
The accuracy of our results degrades beyond the frequency
range in which the response basis was optimized, and energy-optimized bases
for excited states were added explicitly.
Within this range, they provide benchmark values for the scalar
polarizabilities of both states and the polarizability anisotropy of the
$2\,{}^3P_u$ state, at previously unreported accuracy level.

\noindent
{\bf Acknowledgment}
This work was financed by Wroc{\l}aw University of Science and Technology.
The calculations have been carried out in Wroc{\l}aw Center
for Networking and Supercomputing (WCSS, http://wcss.pl).

\end{document}